\documentclass[10pt]{article}
\usepackage{amssymb,amsmath,ae,amsthm}%
\usepackage[dvips]{graphicx,color}%
\usepackage{setspace}
\newcommand{\figdraft}{false}%
\newcommand{\figfile}[1]{#1}%
\theoremstyle{plain}%
\newtheorem{theorem}{Theorem}[]%
\newtheorem{corollary}[theorem]{Corollary}%
\newtheorem{lemma}[theorem]{Lemma}%
\newtheorem{scheme}[theorem]{Scheme}%
\newtheorem{assumption}[theorem]{Assumption}%
\newtheorem{remark}[theorem]{Remark}%
\definecolor{colorGreen}{rgb}{0.,0.67,0}
\definecolor{colorRed}{rgb}{0.67,0.,0}
\definecolor{colorBlue}{rgb}{0.,0.,0.67}

\newcommand{\lapl}{\triangle}

\newcommand{\iu}{\mathtt{i}}
\newcommand{\mhexp}[1]{{{\mathtt{e}}^{#1}}}

\newcommand{\fspace}[1]{{\mathsf{#1}}}
\newcommand{\fspaceL}{\fspace{L}}
\newcommand{\fspaceH}{\fspace{H}}
\newcommand{\fspaceC}{\fspace{C}}

\newcommand{\deq}{{:=}}

\newcommand{\phase}{{\varphi}}

\newcommand{\Rset}{{\mathbb{R}}}
\newcommand{\Zset}{{\mathbb{Z}}}

\newcommand{\Nset}{{\mathbb{N}}}

\newcommand{\cointerval}[2]{[#1,\,#2)}%
\newcommand{\oointerval}[2]{(#1,\,#2)}%
\newcommand{\ccinterval}[2]{[#1,\,#2]}%

\newcommand{\nl}{{\rm nl}}

\newlength{\mhpicDwidth}
\newlength{\mhpicDvsep}
\newlength{\mhpicDhsep}

\newlength{\mhpicPwidth}
\newlength{\mhpicPvsep}
\newlength{\mhpicPhsep}

\newlength{\mhpicWhsep}

\newcommand{\bpair}[2]{{\big({#1},\,{#2}\big)}}

\newcommand{\skp}[2]{{\left\langle{#1},\,{#2}\right\rangle}}
\newcommand{\nskp}[2]{{\langle{#1},\,{#2}\rangle}}
\newcommand{\npair}[2]{{({#1},\,{#2})}}
\newcommand{\at}[1]{{\left({#1}\right)}}
\newcommand{\nat}[1]{(#1)}
\newcommand{\bat}[1]{{\big(#1\big)}}

\newcommand{\ato}[1]{{\left[{#1}\right]}}

\newcommand{\bigpar}{\par\quad\newline\noindent}

\newcommand{\wt}[1]{{\widetilde{#1}}}
\newcommand{\wh}[1]{{\widehat{#1}}}

\newcommand{\norm}[1]{{\|{#1}\|}}
\newcommand{\abs}[1]{\left|{#1}\right|}

\newcommand{\dint}[1]{\,\mathrm{d}#1}

\newcommand{\Ga}{{\Gamma}}

\newcommand{\La}{{\Lambda}}

\newcommand{\ga}{{\gamma}}

\newcommand{\la}{{\lambda}}

\newcommand{\om}{{\omega}}

\newcommand{\calA}{\mathcal{A}}
\newcommand{\calB}{\mathcal{B}}
\newcommand{\calC}{\mathcal{C}}

\newcommand{\calI}{\mathcal{I}}

\newcommand{\calL}{\mathcal{L}}

\newcommand{\calP}{\mathcal{P}}

\newcommand{\calS}{\mathcal{S}}
\newcommand{\calT}{\mathcal{T}}
\newcommand{\calU}{\mathcal{U}}

\usepackage{float}%
\begin{document}
%
%
%
%
%
\title{%
    Periodic travelling waves in convex Klein-Gordon chains
}%
\author{Michael Herrmann\thanks{ %
Oxford Centre for Nonlinear PDE (OxPDE),
Mathematical Institute, 24-29 St Giles', OX1 3LB Oxford, England,
michael.herrmann@maths.ox.ac.uk}}%
\maketitle
%
%
%
%
%
\begin{abstract}
We study Klein-Gordon chains with attractive nearest neighbour
forces and convex on-site potential, and show that there exists a
two-parameter family of periodic travelling waves (wave trains) with
unimodal and even profile functions. Our existence proof is based on
a saddle-point problem with constraints and exploits the invariance
properties of an improvement operator. Finally, we discuss the
numerical computation of wave trains.
\end{abstract}
%
%
%
\quad\newline\noindent%
\begin{minipage}[t]{0.15\textwidth}%
{\small Keywords:} %
\end{minipage}%
\begin{minipage}[t]{0.8\textwidth}%
\small %
\emph{Klein-Gordon chain}, %
\emph{lattice travelling waves}, %
\emph{constrained optimisation} %
\end{minipage}%
\medskip
\newline\noindent
\begin{minipage}[t]{0.15\textwidth}%
\small MSC (2000): %
\end{minipage}%
\begin{minipage}[t]{0.8\textwidth}%
\small %
37K60, 
47J30, 
70F45, 
74J30 
\end{minipage}%
%
%
%
%
%
%
%
%
\section{Introduction}
%

Chains of coupled particles or oscillators have a broad range of
applications in physics, material science, and biology, see
for instance \cite{DPB93,BK04}, and have been studied intensively
during the last decades. For chains of identical
particles coupled by nearest neighbour
interactions the law of motion reads
\begin{align}
\label{Intro:Chain}%
m\ddot{y}_j+\ga\dot{y}_j=%
\Phi^\prime\at{y_{j+1}-y_j}-\Phi^\prime\at{y_j-y_{j-1}}-%
\Psi^\prime\at{y_j}.
\end{align}
Here $y_j\at{t}$ is the displacement of the $j$th particle at time
$t$, $m$ the particle mass, $\ga\geq0$ a damping parameter, and $\Phi$ and $\Psi$ denote
the pair and on-site potential, respectively.  Examples for such chains with $\ga=0$ are
FPU-like chains with $\Psi\equiv0$, and Klein-Gordon chains with
arbitrary on-site but harmonic pair potential. Moreover, in the
case $\Psi\at{y}=\sin{y}$ one either refers to
\eqref{Intro:Chain} as the Frenkel-Kontorova model or the discrete
Sine-Gordon equation.
\bigpar
Major topics in the analysis of atomic chains like
\eqref{Intro:Chain} are the existence and dynamical properties
of coherent structures such as travelling waves and breathers, see
the review article \cite{IJ05}. A travelling
wave is a special solution to \eqref{Intro:Chain} which satisfies
\begin{align}
\label{Intro:Ansatz}%
y_j\at{t}=Y\at{kj-\om{t}},
\end{align}
where $k$ and $\om$ denote the \emph{wave number} and
\emph{frequency}, respectively, and $Y$ is the \emph{profile
function}.
\par
The existence of travelling wave solutions to \eqref{Intro:Chain}
has been investigated by several authors using rather different methods.
For chains with harmonic $\Phi$ we refer to \cite{IK00} which
establishes the existence of small amplitude waves by using spatial
dynamics and centre manifold reduction, see also \cite{IP06}.
Numerical simulations are presented in \cite{DEFW93}, and
\cite{MJKS02} investigates the existence and stability of standing
waves by using a continuum approximation for the small
amplitude limit. More recently, the existence of periodic travelling
waves for the Frenkel-Kontorova model was shown in \cite{Kat05} by means of 
fixed point methods, and
\cite{BZh06} provides existence results in Sine-Gordon chains with
even non-local interactions. Close to our approach are
\cite{KZ08a,KZ08b}, which set the problems also in a variational
framework and prove the existence of supersonic waves for
Sine-Gordon chains. Finally, a lot of literature addresses the
existence of travelling waves in FPU-like chains, compare
\cite{Pan05,IJ05,Her09} and references therein.
\bigpar
In this  paper we restrict ourselves to Klein-Gordon chains with
\emph{convex} on-site potential $\Psi$ and attractive nearest 
neighbour forces, and consider periodic
travelling waves, which in turn are called \emph{wave trains}. 
Due to simple normalisations we can suppose that $m=1$ and 
$\Phi\at{y}=\tfrac{1}{2}y^2$, so the profile of
each wave train must solve the nonlinear advance-delay 
differential equation
\begin{align}
\label{Intro:TW.Eqn0}%
\omega^2{}
\frac{\dint^2}{\dint\phase^2}
Y=\lapl_k{Y}-\Psi^\prime\at{Y},
\end{align}
where $\phase=kj-\om{t}$ abbreviates the
\emph{phase}, and $\lapl_k$ is a discrete
Laplacian with
\begin{align}
\label{Intro:Lapl}
\at{\lapl_k{Y}}\at{\phase}=%
Y\at{\phase+k}+Y\at{\phase-k}-2Y\at{\phase}.
\end{align}
Moreover, since \eqref{Intro:TW.Eqn0} is invariant under shifts in
$\phase$-direction and under the scaling
\begin{align*}
Y\at\phase\rightsquigarrow
Y\at{\la\phase}, \qquad
k\rightsquigarrow\la^{-1}k,
\qquad
\om\rightsquigarrow\la^{-1}\om,
\end{align*}
we can assume that $Y$ is 
$1$-periodic with unit cell $\La=\cointerval{-1/2}{1/2}$.
\bigpar
In order to proof the existence of $1$-periodic solutions to
\eqref{Intro:TW.Eqn0} we follow a variational approach and 
characterise wave trains as solutions to a constraint saddle point
problem for the potential energy. In particular, the frequency $\om$
turns out to be the square root of the Lagrange multiplier, and cannot be prescribed. 
This is different from the standard variational method which 
prescribes the frequency and characterises travelling waves as stationary points of the 
action integral, compare for instance \cite{KZ08a,KZ08b}.
\par
A further key ingredient for our method is an improvement operator, which we
introduce below. Due to the convexity of $\Psi$ this operator possesses nontrivial 
invariant cones, and this allows to establish the existence of wave trains within 
these cones. More precisely, our main result can be stated as follows.
\begin{theorem}
\label{Intro:Result} There exists a two-parameter family of
solutions to \eqref{Intro:TW.Eqn0} such that the profile function
$Y$ is $1$-periodic, unimodal, and even.
\end{theorem}
The paper is organised as follows. In \S\ref{sec:setting} we
describe our variational setting, and \S\ref{sec:Proof} contains
the proof of the main result. Finally, in \S\ref{sec:App} we discuss
the numerical approximation of wave trains and present some
simulations.
%
%
%
%
\section{Variational setting}\label{sec:setting}
%
In the remainder of this paper we rely on the following standing assumption.
\begin{assumption}
\label{Ass:Pot}%
The on-site potential $\Psi:\Rset\to\Rset$ is twice continuously
differentiable and uniformly convex, in the sense that there exist
two constants $0<m<M<\infty$ such that
$m\leq\Phi^{\prime\prime}\at{x}\leq{M}$ for all $x\in\Rset$.
Moreover, $\Psi$ is always normalised by
$\Psi\at{0}=\Psi^\prime\at{0}=0$.
\end{assumption}
This assumption  in particular implies
\begin{align*}
\tfrac{1}{2}mx^2\leq\Psi\at{x}\leq\tfrac{1}{2}Mx^2
,\qquad%
\abs{\Psi^\prime\at{x}}\leq{M}\abs{x}
\end{align*} %
for all $x\in\Rset$, and that $\Psi$ has a unique global minimum at
$x=0$. We mention that the proofs below require $m$ and $M$ to exist
only for all bounded sets, so our results remain valid if
$\Phi^{\prime\prime}$ is bounded and uniformly positive on each
closed interval.
%
%
\paragraph{Spaces and cones of functions}
%
%
We denote by $\fspaceC^k$ the space of all functions that are
periodic with unit cell $\La$ and $k$ times continuously
differentiable, and equip these spaces with their usual norms.
Similarly, $\fspaceL^2$ is the space of all periodic functions which
are square integrable on $\La$, and $\fspaceH^1$ abbreviates the
space of all functions $Y\in\fspaceL^2$ which have a weak derivative
$Y^\prime\in\fspaceL^2$. Both $\fspaceL^2$ and $\fspaceH^1$ are
Hilbert spaces with scalar products
\begin{align*}
\skp{Y_1}{Y_2}_{\fspaceL^2}&
=%
\int_{\La}{Y_1}\at{\phase}{Y_2}\at{\phase}\dint\phase,\qquad
\skp{Y_1}{Y_2}_{\fspaceH^1}
=%
\skp{Y_1^{\prime}}{Y_2^\prime}_{\fspaceL^2}
+\int_{\La}{Y_1}\at{\phase}\dint\phase
\int_{\La}{Y_2}\at{\phase}\dint\phase.
\end{align*}
Note that $\skp{\cdot}{\cdot}_{\fspaceH^1}$ is equivalent to the
standard scalar product as
$Y\mapsto\abs{\int_{\La}Y\at\phase\dint\phase}$ defines a norm for
the constants. Finally, we denote by $\fspaceH^1_0$ the closed
subspace of all functions $X\in\fspaceH^1$ with
$\int_{\La}{X}\at{\phase}\dint\phase=0$, and set
\begin{align*}
\calB_\ga:=\left\{X\in\fspaceH^1_0
\;:\;%
\tfrac{1}{2}\norm{X^\prime}_{2}^2\leq\ga\right\}.
\end{align*}
\bigpar
To characterise the qualitative properties of wave trains we
introduce the cone $\calU$ of all unimodal and even functions on
$\La$, that is
\begin{align*}
\calU&\deq %
\left\{%
Y\in\fspaceC^0\;:\;Y\at{-\phase}
=%
Y\at{\phase}\;\text{and}\;Y\at{\tilde{\phase}}\geq{Y}\at{\phase}\; %
\text{for all}\;0\leq\tilde{\phase}\leq\phase\leq\tfrac{1}{2}
\right\},
\end{align*}
and define a further cone $\calC$ by
\begin{align*}
\calC\deq\left\{Y\in\fspace{C}^2\;:\;-Y^{\prime\prime}\in\calU\right\}.
\end{align*}
By construction, we have $Y\in\calU$ if and only if
$-Y\at{\tfrac{1}{2}+\cdot}\in\calU$, and below in Lemma
\ref{Lem:UsubsetC} we show that $\calC$ is a subcone of $\calU$.
%
%
%
%
\paragraph{Examples of wave trains}
%
%
We proceed with some explicit solutions 
to the wave train equation \eqref{Intro:TW.Eqn0}.
\begin{remark}
%
For $k=0$ the wave train equation reduces to
the oscillator ODE
\begin{align}
\label{Eqn:WaveTrainODE}
\omega^2{}Y^{\prime\prime}=-\Psi^\prime\at{Y}.
\end{align}
Therefore, there exists a one-parameter family of wave trains with
$Y\in\calC$, which can be parametrised by the energy
$E=\tfrac{1}{2}\om^2{Y^\prime}^2+\Psi\at{Y}$.
\end{remark}
\begin{proof}
For given $E>0$ we pick $y<0$ such that $\Psi\at{y}=E$ and define
the function $\wt{Y}=\wt{Y}\at{t}$ with $t\in\Rset$ as solution to
$\wt{Y}^{\prime\prime}\at{t}=-\Psi\bat{\wt{Y}\at{t}}$ with initial
conditions $\wt{Y}\at{0}=y$ and $\wt{Y}^\prime\at{0}=0$. According
to the properties of $\Psi$, compare Assumption \ref{Ass:Pot}, this
function $\wt{Y}$ is periodic with period $T=T\at{E}$, satisfies
$\wt{Y}\at{t}=\wt{Y}\at{T-t}$, and is strictly increasing on
$\ccinterval{0}{T/2}$. We now define $Y\in\calU$ by
$Y\at{\phase}=\wt{Y}\bat{T\phase+T/2}$, which solves
\eqref{Eqn:WaveTrainODE} with $\om=1/T$. %
\end{proof}
\begin{remark}
%
For the harmonic potential $\Psi\at{x}=\tfrac{c}{2}{x}^2$
there exists a two-parameter family of wave trains
with $Y\in\calC$ given by
\begin{align}
\label{Ex:Harm.Eqn1}
{Y}\at\phase=a\cos\at{2\pi\phase},\quad\om^2=\frac{4\sin\at{k\pi}^2+c}{4\pi^2},
\end{align}
where the wave number $k$ and the amplitude $a>0$ are the free
parameters.
\end{remark}
\begin{proof}
We introduce the Fourier transform $\at{y_m}_{m\in\Zset}$ of $Y$, that means
\begin{align*}
Y\at{\phase}=\sum\limits_{m\in\Zset}y_m\mhexp{\iu2\pi{m}\phase}
,\qquad %
y_{m}=\int_{\La}Y\at\phase\mhexp{-\iu2\pi{m}\phase}\dint\phase,
\end{align*}
and find \eqref{Intro:TW.Eqn0} to be equivalent to
\begin{align*}
4\pi^2m^2\omega^2{}y_{m}=\at{2\at{1-\cos\at{2\pi{mk}}}+c}y_{m}
,\quad%
{m}\in\Zset.
\end{align*}
To solve this we set $y_m=y_{-m}=\tfrac{1}{2}\,\delta_m^1$,
with $\delta_m^1$ being the Kronecker delta, and thanks to
$1-\cos{2z}=2\sin^2{z}$ we obtain \eqref{Ex:Harm.Eqn1}.
\end{proof}
%
%
%
%
\paragraph{The Lagrangian structure}
%
%
For general $\Psi$ and $k\neq0$ we cannot solve the wave train
equation explicitly but need more sophisticated arguments to prove
the existence of solutions. The starting point for each variational
approach is the \emph{Lagrangian} of a wave train 
\begin{align*}
\calL\at{Y}=\om^2\Ga\at{Y}-\calP_k\at{Y},\qquad
\Ga\at{Y}=\tfrac{1}{2}\int_{\La}{Y}^\prime\at{\phase}^2\dint\phase,
\end{align*}
with kinetic energy $\om^2\Ga\at{Y}$ and potential energy 
\begin{align*}
\calP_k\at{Y}=\tfrac{1}{2}\norm{\nabla_k{Y}}_2^2+\calP_\nl\at{Y}
,\qquad%
\calP_\nl\at{Y}=\int_{\La}\Psi\at{Y\at\phase}\dint\phase,
\end{align*}
where ${\nabla_k{Y}}\at\phase=Y\at{\phase+k/2}-Y\at{\phase-k/2}$.
Notice that $\calP_k$ is convex due to Assumption
\ref{Ass:Pot}, and that the discrete difference operator $\nabla_k$
satisfies
\begin{align}
\label{Intro:PropsNabla}%
{\lapl_k}=\nabla_k\nabla_k,\qquad\nabla_k^\ast=-\nabla_k
\end{align}
with $\lapl_k$ as in \eqref{Intro:Lapl} and $\ast$ denoting the
$\fspaceL^2$-adjoint. 
\bigpar
Our variational method relies on the following main observation. 
Suppose $Y\in\fspaceH^1$ with $\tfrac{1}{2}\norm{Y}_{\fspaceH^1}^2=\ga>0$ 
is a wave train. Then \eqref{Intro:TW.Eqn0} implies that $Y$ is a stationary
point of $\calP_k$ under the constraint $\Ga\leq\ga$, where $\om^2$ plays 
the role of an Lagrange multiplier. To clarify this stationarity 
condition, we write $Y=x+X$ with $x\in\Rset$ and $X\in\fspaceH^1_0$, and 
restate the wave train equation as
\begin{align}
\label{Intro:TW.Eqn}%
\omega^2{}X^{\prime\prime}=\lapl_k{X}-\Psi^\prime\at{x+X}.
\end{align}
The convexity of $\calP_k$ now implies that each wave train $Y=x+X$
is a minimiser for $\calP_k$ with respect to unconstrained
variations of $x$. With respect to variations of $X\in\calB_\ga$,
however, the only minimiser of $\calP_k$ in $\calB_\ga$ is the
trivial solution $X=0$ with multiplier $\om^2=0$, and hence we are
interested in other types of stationary points. Below we show 
that there exist saddle point solutions $Y$ to
\eqref{Intro:TW.Eqn0} which posses a positive multiplier $\om^2>0$
as they correspond to a maximiser of $\calP_k$ with respect to
variations of $X$. Moreover, due to the properties of the 
aforementioned improvement operator we can additionally impose the 
condition $X\in\calC$, and hence we substantiate
Theorem \ref{Intro:Result} as follows.
\begin{theorem}
\label{Theo:Main} %
For given $\ga>0$ and $k\in{\Lambda}$ there exists a pair
$\npair{\wh{X}}{\wh{x}}\in\calC\cap\calB_\ga\times\Rset$ such that
\begin{align*}
\calP_k\nat{\wh{x}+\wh{X}}
&=%
\max\limits_{X\in\calC\cap\calB_\ga}
\min\limits_{x\in\Rset}\calP_k\at{x+X}.
\end{align*}
The function $\wh{Y}=\wh{x}+\wh{X}\in\calC$ is then a wave train,
that means it solves \eqref{Intro:TW.Eqn0} for some $\om^2>0$.
\end{theorem}
%
%
%
%
\section{Proof of the Existence Result}\label{sec:Proof}%
%
%
In this section we always suppose that the parameters $\ga>0$ and
$k\in{\Lambda}$ are arbitrary but fixed.
\begin{remark}
\label{Rem:SpacesEmbedding}%
$\fspaceH^1_0$ is compactly embedded in $\fspaceC^0$ and
$\fspaceL^2$ with
$\norm{X}_2\leq\norm{X}_\infty\leq\norm{X^\prime}_2$ for all
$X\in\fspaceH^1_0$. In particular, weak convergence in
$\fspaceH^1_0$ implies strong convergence in both $\fspaceC^0$ and
$\fspaceL^2$.
\end{remark}
\begin{proof}
For given $X\in\fspaceH^1_0$ and arbitrary $\phase_0$, $\phase_1$
the integral representation
\begin{math}
X\at{\phase_1}-X\at{\phase_0}=
\int_{\phase_0}^{\phase_1}{X}^\prime%
\at{\phase}\dint\phase
\end{math} %
and H\"olders inequality imply
\begin{math}
-\norm{X^\prime}_2\leq{X}\at{\phase_1}-X\at{\phase_0}\leq\norm{X^\prime}_2,
\end{math} %
where we used that $\int_\La\dint\phase=1$. Integrating these
estimates with respect to $\phase_0\in\Lambda$ gives
$\norm{X}_\infty\leq\norm{X^\prime}_2$, and
$\norm{X}_2\leq\norm{X}_\infty$ follows again from H\"olders
inequality.
\end{proof}
%
%
%
%
\paragraph*{Properties of the potential energy}%
%
%
We proceed with some elementary properties of $\calP_k$.
\begin{lemma}
\label{Lem:PropsFunctP}%
The functional $\calP_{k}:\fspaceL^2\to\Rset$ has the following
properties:
\begin{enumerate}
\item
It is well-defined, nonnegative, continuous, and strictly convex.
\item
It is G\^{a}teaux- differentiable with
$\partial_Y\calP_k\ato{Y}=-\lapl_k{Y}+\Psi^\prime\at{Y}$, and its
derivative $\partial_Y\calP_k:\fspaceL^2\to\fspaceL^2$ is a
continuous operator.
\item
For all $Y_1,\,Y_2\in\fspaceL^2$ we have
\begin{align}
\label{Lem:PropsFunctP.Eqn1} %
\calP_k\at{Y_2}-\calP_k\at{Y_1}
\geq&
\tfrac{m}{2}\norm{Y_2-Y_1}_2^2+%
\skp{\partial_Y\calP\ato{Y_1}}{Y_2-Y_1}_{\fspaceL^2}.
\end{align}
\item
$Y\neq0$ implies $\calP_k\at{Y}>0$ and
$\partial_Y\calP_k\ato{Y}\neq0$.
\end{enumerate}
\end{lemma}
\begin{proof}
The proof of the first two assertions is straight forward. Towards
\eqref{Lem:PropsFunctP.Eqn1} we notice that \eqref{Intro:PropsNabla}
implies
\begin{align*}
\norm{\nabla_k{Y}_1}_2^2=-\skp{\lapl_k{Y}_1}{{Y}_1}_{\fspaceL^2},
\qquad
\tfrac{1}{2}\norm{\nabla_k{Y}_2}_2^2+
\tfrac{1}{2}\norm{\nabla_k{Y}_1}_2^2
\geq%
-\skp{\lapl_k{Y}_1}{Y_2}_{\fspaceL^2},
\end{align*}
and hence
\begin{align}
\label{Lem:PropsFunctP.Eqn2a} %
\tfrac{1}{2}\norm{\nabla_k{Y}_2}_2^2-
\tfrac{1}{2}\norm{\nabla_k{Y}_1}_2^2
\geq%
-\skp{\lapl_k{Y}_1}{Y_2-Y_1}_{\fspaceL^2}.
\end{align}
Moreover, the convexity inequality for $\Psi$ provides
\begin{align*}
\Psi\at{y_2}-\Psi\at{y_1}
\geq%
\tfrac{m}{2}\at{y_2-y_1}^2+\Psi^{\prime}\at{y_1}\at{y_2-y_1},\qquad
y_1,y_2\in\Rset.
\end{align*}
We set $y_i=Y_i\at\phase$ and integrate
this identity with respect to $\phase\in\La$ to obtain
\begin{align}
\label{Lem:PropsFunctP.Eqn2b}
\calP_\nl\at{Y_2}-\calP_\nl\at{Y_1}
\geq%
\tfrac{m}{2}\norm{Y_2-Y_1}_2^2+
\skp{\partial_Y\calP_\nl\ato{Y_1}}{Y_2-Y_1}_{\fspaceL^2}.
\end{align}
The estimate \eqref{Lem:PropsFunctP.Eqn1} then follows by adding
\eqref{Lem:PropsFunctP.Eqn2a} and \eqref{Lem:PropsFunctP.Eqn2b}. In
particular, exploiting \eqref{Lem:PropsFunctP.Eqn1} with $Y_2=Y$ and
$Y_1=0$ we find  $\calP_k\at{Y}>0$ for all $Y\neq0$. To complete the
proof we suppose that $\partial_Y\calP_k\ato{Y}=0$. Then
\eqref{Lem:PropsFunctP.Eqn1} with $Y_2=0$ and $Y_1=Y$ gives
\begin{align*}
0\geq-\calP_k\at{Y}
=%
\calP_k\at{0}-\calP_k\at{Y}
\geq%
\tfrac{m}{2}\norm{Y}_2^2,
\end{align*}
and hence $Y=0$.
\end{proof}
Next we show that for each $X$ we can choose a unique $x$ by
minimising the potential energy.
\begin{lemma}
\label{Lem:MinInSaddle} %
There exists a unique and continuous map
$\wh{x}:\fspaceH^1_0\mapsto\Rset$ such that
\begin{align}
\label{Lem:MinInSaddle.Eqn1}%
\int_{\Lambda}\Psi^\prime\at{\wh{x}\at{X}+X\at\phase}\dint\phase
=0,%
\qquad%
\calP_k\at{\wh{x}\at{X}+X}
=%
\min\limits_{x\in\Rset}\calP_k\at{x+X}
\end{align}
for all $X\in\fspaceH^1_0$.
\end{lemma}
\begin{proof}
By assumption \ref{Ass:Pot}, the function
\begin{math}
\psi\at{x}=\int_{\Lambda}\Psi\at{x+X\at\phase}\dint\phase
\end{math} %
is well-defined and twice continuously differentiable with
derivatives
\begin{math}
\psi^{\at{i}}\at{x}
=%
\int_{\Lambda}\Psi^{\at{i}}\at{x+X\at\phase}\dint\phase
\end{math} %
for ${i}=1,\,2$. In particular, $\psi$ is uniformly convex with
$M\geq\psi^{\prime\prime}\at{x}\geq{m}>0$, and hence there exists a
unique minimiser $\wh{x}=\wh{x}\at{X}$ with
$\psi^\prime\at{\wh{x}}=0$. It is straight forward that
$\wh{x}\at{X}$ depends continuously on $X$ with respect to the
strong topology in $\fspaceC^0$, and the compact embedding from
Remark \ref{Rem:SpacesEmbedding} implies the continuity with respect
to the weak topology in $\fspaceH^1_0$.
\end{proof}
In what follows we consider the reduced potential energy functional
\begin{align*}
\wh{\calP}_k:\fspaceH^1_0\to\Rset,
\qquad\wh{\calP}_k\at{X}&\deq{\calP}_k\at{\wh{x}\at{X}+X},
\end{align*}
and aim to show that there exists maximisers for $\wh{\calP}_k$ in
$\calC\cap\calB_\ga$. To this end we draw the following conclusion 
from Lemma \ref{Lem:MinInSaddle}.
\begin{remark}
\label{Rem:PropsEffEnergy}
The functional $\wh{\calP}_k$ is weakly continuous on $\fspaceH^1_0$.
\end{remark}
%
%
%
\paragraph*{The improvement operator}
%
%
As a main ingredient for the proof of Theorem \ref{Theo:Main} we
introduce the \emph{improvement operator} $\calT_{k,\,\ga}$ as
follows. For each $X\neq0$ the function
$\tilde{X}\deq\calT_{k,\,\ga}\ato{X}$ satisfies
\begin{align}
\label{ImprovOp.Eqn1}%
{\om}^2\tilde{X}^{\prime\prime}
&=%
\lapl_k{X}-\Psi^{\prime}\at{\wh{x}\at{X}+X},
\end{align}
where ${\om}^2$ is chosen such that $\tilde{X}\in\partial\calB_\ga$.
Notice that $\tilde{X}$ is well-defined as
\eqref{Lem:MinInSaddle.Eqn1}$_1$ implies that the right hand side in
\eqref{ImprovOp.Eqn1} vanishes when integrating over $\La$. We
further define the integral operator
\begin{align}
\label{Eqn:DefOpI}
\at{\calI{X}}\at\phase
&\deq%
\int\limits_0^\phase{X}\at{\tilde\phase}\dint\tilde\phase-
\int\limits_{\La}\int\limits_0^{\bar\phase}{X}\at{\tilde\phase}
\dint\tilde\phase\dint\bar\phase,
\end{align}
and thanks to $\at{\calI{X}}^\prime=X$ we rewrite
\eqref{ImprovOp.Eqn1} as
\begin{align}
\label{ImprovOp.Eqn2}%
\calT_{k,\,\ga}\ato{X}\deq-\frac{\calI\calI{Z}}{\om^2}
,\qquad%
\om^2\deq\frac{\norm{\calI{Z}}_2}{\sqrt{2\ga}}
,\qquad%
{Z}\deq\partial_Y\calP_k\ato{\wh{x}\at{X}+X}=-\lapl_k{X}+
\Psi^\prime\at{\wh{x}\at{X}+X}.
\end{align}
\begin{remark}
\label{Rem:PropsOpI}%
The operator $\calI$ is a well-defined and a compact endomorphism of
$\fspaceH^1_0$. In particular, $X_n\to{X}_\infty$ weakly implies
$\calI{X}_n\to\calI{X}_\infty$ strongly, and $\calI\at{X}=0$ implies
$X=0$.
\end{remark}
\begin{lemma}
\label{Lem:PropsImprovOp}%
The improvement operator maps $\calB_\ga\setminus\{0\}$ to
$\partial\calB_\ga\setminus\{0\}$ and weakly convergent sequences to
strongly convergent ones. Moreover, it satisfies
\begin{align}
\label{Lem:PropsImprovOp.Eqn1}
\wh{\calP}_k\nat{\calT_{k,\,\ga}\ato{X}}\geq\wh{\calP}_k\at{X},
\end{align}
where equality holds if and only if $X=\calT_{k,\,\ga}\ato{X}$.
\end{lemma}
\begin{proof}
According to \eqref{Eqn:DefOpI} and \eqref{ImprovOp.Eqn2}, the
function $\calT_{k,\,\ga}\ato{X}\in\partial\calB_\ga$ is 
well-defined as long as $\calI{Z}\neq0$, which holds true if and only if
$X\neq0$, compare Remark \ref{Rem:PropsOpI} and Lemma
\ref{Lem:PropsFunctP}. Moreover, the claimed convergence properties
are implied by \eqref{ImprovOp.Eqn2} and Remark \ref{Rem:PropsOpI}.
Now let $X\in\calB_\ga$ be fixed, and set $\wt{X}=\calT_{k,\,\ga}\ato{X}$ and
$x=\wh{x}\at{X}$, $\wt{x}=\wh{x}\nat{\wt{X}}$. Then
\eqref{ImprovOp.Eqn1} reads
$\om^2\tilde{X}^{\prime\prime}=-\partial_Y\calP_k\at{x+X}$, and from
\eqref{Lem:PropsFunctP.Eqn1} we infer that
\begin{align*}
\calP_k\nat{\wt{x}+\wt{X}}-\calP_k\at{x+X}
\geq%
{\om^2}%
\nskp{-\wt{X}^{\prime\prime}}%
{\wt{x}-x+\wt{X}-X}_{\fspaceL^2}=
{\om^2}%
\nskp{\wt{X}}{\wt{X}-X}_{\fspaceH^1_0}.
\end{align*}
This implies \eqref{Lem:PropsImprovOp.Eqn1} due to
$2\ga=\norm{\wt{X}}_{\fspaceH^1_0}^2$ and
\begin{math}
\nskp{\wt{X}}{X}_{\fspaceH^1_0}
\leq
\norm{\wt{X}}_{\fspaceH^1_0}\norm{X}_{\fspaceH^1_0}\leq2\ga.
\end{math} %
Moreover, we have equality in \eqref{Lem:PropsImprovOp.Eqn1}
if and only if
\begin{math}
\nskp{\wt{X}}{X}_{\fspaceH^1_0}
=
\norm{\wt{X}}_{\fspaceH^1_0}^2=\norm{X}_{\fspaceH^1_0}^2,
\end{math} %
that means $X=\wt{X}$.
\end{proof}
The following implication of Lemma \ref{Lem:PropsImprovOp} is key
for our existence proof.
\begin{corollary}
\label{Corr:ExistWaveTrains}%
Suppose that the set $\calS\subset\fspaceL^2$ is invariant under the
action of $\calT_{k,\,\ga}$. Then, each maximiser $X$ for
$\wh\calP_k$ in $\calS\cap\calB_\ga$ is a wave train with
non-vanishing frequency and satisfies
$\tfrac{1}{2}\norm{X}_2^2=\ga$.
\end{corollary}
\begin{proof}
Let $X$ be a maximiser for $\wh\calP_k$ in $\calS\cap\calB_\ga$, and
recall that $X\neq0$ according to Lemma \ref{Lem:PropsFunctP}. By
assumption,  $\wt{X}\deq\calT_{k,\,\ga}\ato{X}$  satisfies
$\wt{X}\in\calS\cap\partial\calB_\ga$ as well as
$\wh\calP_k\at{X}\geq\wh\calP_k\nat{\wt{X}}$, so Lemma
\ref{Lem:PropsImprovOp} provides
$\wh\calP_k\at{X}=\wh\calP_k\nat{\wt{X}}$, and hence $X=\wt{X}$.
\end{proof}
Recall that \eqref{Intro:TW.Eqn} can be viewed as the Euler-Lagrange
equation for the optimisation problem
$\wh{\calP}_k\nat{{X}}\to\max{}$ with ${X}\in\calB_\ga$, where the
Lagrange multiplier corresponding to $\int_\La{X}\at\phase=0$
vanishes due to the choice of $\wh{x}\at{X}$. In particular, the
fact that each maximiser for $\wh{\calP}_k$ in the smaller set
$\calS\cap\calB_\ga$ satisfies \eqref{Intro:TW.Eqn} without further
multipliers is not clear a priori but provided by the invariance of
$\calS$ under the action of $\calT_{k,\,\ga}$.
%
%
\paragraph*{Properties of the cones $\calC$ and $\calU$}%
%
%
Our next result is rather elementary but provides an important
building block for our existence result.
\begin{lemma}
\label{Lem:UsubsetC} %
$X\in\calC$ implies both $X\in\calU$ and $-\lapl_k{X}\in\calU$.
\end{lemma}
\begin{proof}
Thanks to $Z\deq-X^{\prime\prime}\in\calU$ the function $X$ is even
and there exists $\phase_0\in\oointerval{0}{1/2}$ such that
$Z\at\phase\geq0$ for all $0\leq\phase\leq\phase_0$ but
$Z\at\phase\leq0$ for all $\phase_0\leq\phase\leq{1/2}$. Therefore,
$X$ is concave on $\ccinterval{-\phase_0}{\phase_0}$ with maximum in
$0$, and we infer that
${X}^\prime\at{-\phase}\geq{X}^\prime\at{0}=0\geq{X}^\prime\at{\phase}$
for all $0\leq\phase\leq\phase_0$. Similarly, ${X}$ is convex in
$\ccinterval{\phase_0}{1-\phase_0}$ with minimum in $1/2$ and thus
we have shown $X\in\calU$.
\par
Towards the second claim we introduce the averaging operator
\begin{align*}
\at{\calA_k{X}}\at\phase
=%
\int\limits_{\phase-k/2}^{\phase+k/2}{X}\at{\tilde\phase}
\dint{\tilde\phase},
\end{align*}
and a direct computation shows
\begin{math}
\int_\La\at{\calA_k{X}\at\phase}\dint\phase
=%
k\int_\La{X}\at\phase\dint\phase
\end{math} %
and that $\calA_k$ is even provided that $X$ is even. Moreover, we
have $\nabla_k{X}=\calA_k{X^\prime}=\at{\calA_k{X}}^\prime$ and
hence $\lapl_k{X}=\calA_k^2{X^{\prime\prime}}$ for all $X$. It
remains to show that $\calA_k{Z}$ is non-increasing on
$\phase\in\ccinterval{0}{1/2}$ for all $Z\in\calU$, and to this end
we discuss the following cases for $0\leq\phase\leq{1}/2$:
$\at{a}$ $\phase-k/2\leq0\leq\phase+k/2\leq1/2$, %
$\at{b}$ $\phase-k/2\leq0\leq1/2\leq\phase+k/2$, %
$\at{c}$ $0\leq\phase-k/2\leq\phase+k/2\leq1/2$, %
$\at{d}$ $0\leq\phase-k/2\leq1/2\leq\phase+k/2$. %
\par
In view of  $Z\at\phase=Z\at{-\phase}$ we find for case $\at{a}$ the
estimate
$\at{\calA_k{Z}}^\prime\at\phase=Z\at{k/2+\phase}-Z\at{k/2-\phase}<0$,
and $Z\at\phase=Z\at{1-\phase}$ provides for case $\at{b}$ that
$\at{\calA_k{Z}}^\prime\at\phase=Z\at{1-k/2-\phase}-Z\at{k/2-\phase}<0$.
Similarly, for case $\at{c}$ and $\at{d}$ we obtain
$\at{\calA_k{Z}}^\prime\at\phase=Z\at{\phase+k/2}-Z\at{\phase-k/2}<0$
and
$\at{\calA_k{Z}}^\prime\at\phase=Z\at{1-k/2-\phase}-Z\at{\phase-k/2}<0$,
respectively, and the proof is finished.
\end{proof}
\begin{corollary}
$X\in\calC$ implies $\calT_{k,\,\ga}\ato{X}\in\calC$.
\end{corollary}
\begin{proof}
For each $X\in\calC$ Lemma \ref{Lem:UsubsetC} provides
$-\lapl_kX\in\calU$ and $X\in\calU$, and from the latter we conclude
that $\Psi^\prime\at{\wh{x}\at{X}+X}\in\calU$ as $\Psi^\prime$ is
monotonically increasing. Consequently, the right hand side in
\eqref{ImprovOp.Eqn1} is contained in $-\calU$, and this implies
$\calT_{k,\,\ga}\ato{X}\in\calC$.
\end{proof}
%
%
\paragraph*{Existence of maximisers}%
%
%
The cone $\calC$ is not closed under weak convergence in
$\fspaceH^1_0$, but we can prove the following result.
\begin{lemma}
\label{Lem:ImprovOp.Compactness}
Let $X_n\subset\calC\cap\calB_\ga$ be any sequence with
$X_n\to{X}_\infty$ weakly in $\fspaceH^1_0$ for some limit
$X_\infty\neq0$. Then we have
$\calT_{k,\,\ga}\ato{X_n}\to\calT_{k,\,\ga}\ato{X_\infty}$ strongly
in $\fspaceH^1_0$ and
$\calT_{k,\,\ga}\ato{X_\infty}\in\calC\cap\partial\calB_\ga$.
\end{lemma}
\begin{proof}
For all $n\in\Nset\cup\{\infty\}$ let
\begin{align*}
\wt{X}_n\deq\calT_{k,\,\ga}\ato{X_n}
,\quad%
x_n\deq\wh{x}\at{X_n}
,\quad%
Z_n\deq\partial_Y\calP\at{x_n+X_n}
,\quad%
\om_n^2\deq \frac{\norm{\calI{Z}_n}_2}{\sqrt{2\ga}},
\end{align*}
and recall that \eqref{ImprovOp.Eqn2} implies
\begin{align*}
\om_n^2\wt{X}_n^{\prime\prime}
=%
-Z_n
=%
\lapl_k{X}_n-\Psi^\prime\at{x_n+X_n}.
\end{align*}
Remark \ref{Rem:SpacesEmbedding} and Lemma \ref{Lem:PropsFunctP}
provide $Z_n\to{Z}_\infty$ strongly in $\fspaceC^0$, and hence also
$\om_n\to\om_\infty$. Moreover, we have $\om_\infty\neq0$ since
$X_\infty\neq0$ implies $Z_\infty\neq0$ and $\calI{Z}_\infty\neq0$,
compare  Remark \ref{Rem:PropsOpI} and Lemma \ref{Lem:PropsFunctP}.
We conclude that
$\wt{X}_n^{\prime\prime}\to\wt{X}_\infty^{\prime\prime}$ strongly in
$\fspaceC^0$, thus $\wt{X}_\infty\in\calB_\ga$, and since $\calU$ is
closed we infer that $Z_\infty\in\calU$, and therefore
$\wt{X}_\infty\in\calC$.
\end{proof}
\begin{corollary}
\label{Corr:ExistMax} %
$\wh{\calP}_k$ attains its maximum in $\calC\cap\calB_\ga$.
\end{corollary}
\begin{proof}
Suppose that $\at{X_n}_{n\in\Nset}\subset\calC\cap\calB_\ga$ is a
maximising sequence for $\wh{\calP}_k$. By weak compactness we can
extract a (not relabelled) subsequence such that $X_n\to{X}_\infty$
weakly in $\fspaceH^1_0$ for some limit $X_\infty$ and Remark
\ref{Rem:PropsEffEnergy} implies
$\wh{\calP}_k\at{X_\infty}=\sup\wh{\calP}|_{\calC\cap\calB_\ga}$, 
and hence $X_\infty\neq0$. Moreover, according to Lemma
\ref{Lem:ImprovOp.Compactness} we have
$\wt{X}_\infty\deq\calT_{k,\,\ga}\ato{X_\infty}\in\partial\calB_\ga\cap\calC$,
and from \eqref{Lem:PropsImprovOp.Eqn1} we infer that
$\wt{X}_\infty$ is a maximiser.
\end{proof}
The combination of Corollary \ref{Corr:ExistWaveTrains} and
Corollary \ref{Corr:ExistMax} gives the proof of Theorem
\ref{Theo:Main}.
%
%
%
%
\section{Approximation of Wave Trains}\label{sec:App}
%
By view of the preceding results it seems natural to approximate
wave trains by the following abstract iteration scheme for fixed
points of $\calT_{k,\,\ga}$.
\begin{scheme}
\label{Scheme} %
For given parameters $\ga>0$, $k\in\La$ and fixed initial value
$X_0\in\calC\cap\partial\calB_\ga$ with $X_0\neq0$ we define
sequences
\begin{align*}
\at{x_i}_i\subset\Rset,\qquad\at{\om_i^2}_i\subset\Rset,
\qquad\at{X_i}_i\subset\calC\cap\partial\calB_\ga
\end{align*}
by the following recursion:
\begin{enumerate}
\item solve the scalar optimisation problem for $x_i=\wh{x}\at{X_i}$,
\item compute $Z_i=-\lapl_k{X}_i+\Psi^\prime\at{x_i+X_i}$,
\item solve ${U}^{\prime\prime}_i=-Z_i$ for ${U}_{i}\in\fspaceH^1_0$,
\item compute $\om_i^2=\norm{U^\prime_i}_{\fspaceL^2}/\sqrt{2\ga}$,
\item set $X_{i+1}=U_i/\om_i^2$.
\end{enumerate}
\end{scheme}
\begin{par}
From a mathematical point of view the account of this scheme is
limited for the following reasons: $\at{i}$ We have no convergence
proof. $\at{ii}$ Due to the lack of uniqueness results it is not
clear whether or not Corollary \ref{Corr:ExistWaveTrains} covers all
fixed points of $\calT_{k,\,\ga}$ in $\calC\cap\calB_\ga$. Nevertheless, suitable discrete
variants are easily derived and work very well in numerical
simulations.
\end{par}
\begin{figure}[ht!]
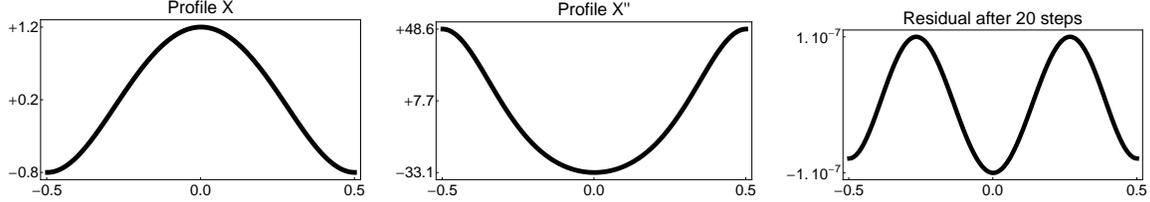
%
  \centering{%
  \includegraphics[width=0.30\textwidth, draft=\figdraft]%
  {\figfile{ex1_profile_x}}%
  \hspace{0.025\textwidth}%
  \includegraphics[width=0.30\textwidth, draft=\figdraft]%
  {\figfile{ex1_profile_xpp}}%
  \hspace{0.025\textwidth}%
  \includegraphics[width=0.30\textwidth, draft=\figdraft]%
  {\figfile{ex1_residual}}%
  }%
  \caption{%
        Profile $X$ with second derivative and
        residual for the data from \eqref{data.Ex1}.
  }%
  \label{NumFig:Ex1}%
\end{figure}%
\begin{par}
In a simple discrete analog to  Scheme \ref{Scheme} each profile
function $X_i$ is identified with a vector $\nat{X_i^j}_{j=1..N}$
via $X^j_i={X_i\at{-1/2+j/N}}$, where $k$ is supposed to be a
multiple of $1/N$, and the derivative $X_i^\prime$ is approximated
by centred finite differences. Moreover, integrals with respect to
$\phase$ are replaced by their Riemann sums and $x_i$ is computed by
a discrete gradient flow for the function
$x\mapsto\sum_{j=1..N}\Psi\nat{x+X_i^j}$. This numerical approach is
illustrated in Figure \ref{NumFig:Ex1} for the data
\begin{align}
\label{data.Ex1}%
\ga=10,\qquad%
k=0.1,\qquad\Psi^{\prime\prime}\at{x}=\exp\at{-x},
\end{align}
with $N=800$ and
$X_0\at\phase=\sqrt{\ga}\pi^{-1}\cos\at{2\pi\phase}$. Under the
iteration the functions $X_i$ converge to a wave train, that means a
fixed point of $\calT_\ga$, with unimodal and even second
derivative. Moreover, numerical simulations indicate that the limit
profile is independent of the initial profile $X_0\in\calC$.
\end{par}
\begin{figure}[ht!]
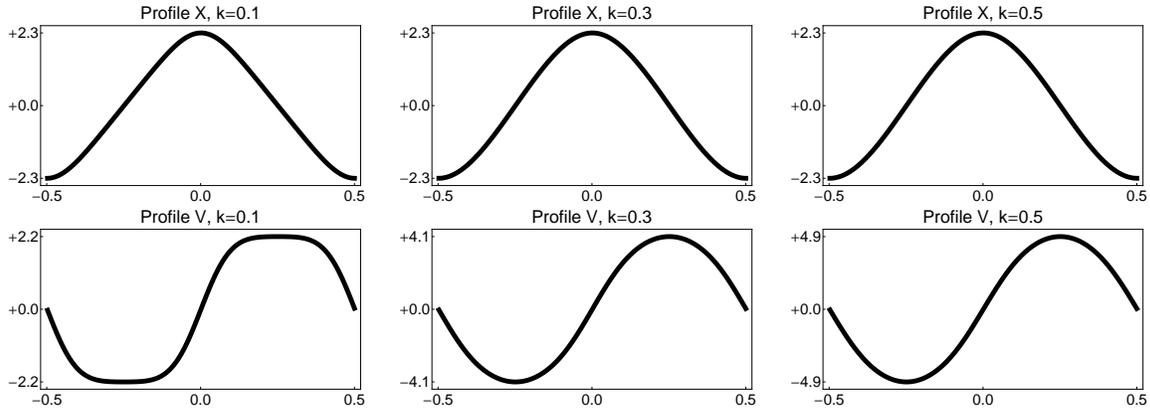
%
  \centering{%
  \includegraphics[width=0.30\textwidth, draft=\figdraft]%
  {\figfile{ex3_profile_x_1}}%
  \hspace{0.025\textwidth}%
  \includegraphics[width=0.30\textwidth, draft=\figdraft]%
  {\figfile{ex3_profile_x_2}}%
  \hspace{0.025\textwidth}%
  \includegraphics[width=0.30\textwidth, draft=\figdraft]%
  {\figfile{ex3_profile_x_3}}%
  \\%
  \includegraphics[width=0.30\textwidth, draft=\figdraft]%
  {\figfile{ex3_profile_v_1}}%
  \hspace{0.025\textwidth}%
  \includegraphics[width=0.30\textwidth, draft=\figdraft]%
  {\figfile{ex3_profile_v_2}}%
  \hspace{0.025\textwidth}%
  \includegraphics[width=0.30\textwidth, draft=\figdraft]%
  {\figfile{ex3_profile_v_3}}%
  }%
  \caption{%
        Profiles $X$ and $V$ for the data from
        \eqref{data.Ex2}.
  }%
  \label{NumFig:Ex2}%
\end{figure}%
\begin{par}
Further numerical results are shown in Figure
\ref{NumFig:Ex2} and correspond to
\begin{align}
\label{data.Ex2}%
\ga=50.,\qquad{k}\in\{0.1,\,0.3,\,0.5\},\qquad%
\Psi^{\prime\prime}\at{x}=1+x^2,
\end{align}
where $N$ and $X_0$ are chosen as above. Here we plot additionally
the profile function $V=-\om{X^\prime}$, which describes the atomic
velocities in a wave train via $\dot{y_j}\at{t}={V}\at{kj-\om{t}}$,
compare \eqref{Intro:Ansatz}.
\end{par}
\begin{figure}[ht!]
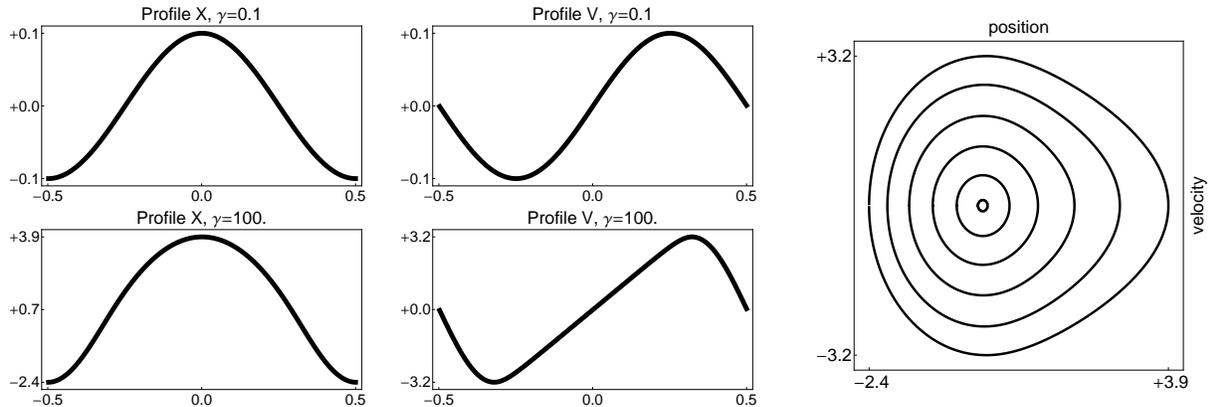
%
\centering{%
    \begin{minipage}[c]{0.3\textwidth}%
        \includegraphics[width=\textwidth, draft=\figdraft]%
        {\figfile{ex2_profile_x_1}}%
        \\%
        \includegraphics[width=\textwidth, draft=\figdraft]%
        {\figfile{ex2_profile_x_2}}%
    \end{minipage}%
    \hspace{0.025\textwidth}%
    \begin{minipage}[c]{0.3\textwidth}%
        \includegraphics[width=\textwidth, draft=\figdraft]%
        {\figfile{ex2_profile_v_1}}%
        \\%
        \includegraphics[width=\textwidth, draft=\figdraft]%
        {\figfile{ex2_profile_v_2}}%
    \end{minipage}%
    \hspace{0.025\textwidth}%
    \begin{minipage}[c]{0.35\textwidth}%
        \vspace{0.05\textwidth}%
        \includegraphics[width=\textwidth, draft=\figdraft]%
        {\figfile{ex2_traces}}%
    \end{minipage}%
  }%
  \caption{%
        Simulations for the data from \eqref{data.Ex3}:
        Profiles $X$ and $V$ for minimal and maximal value of $\ga$,
        and superposition of six traces of wave trains.
  }%
  \label{NumFig:Ex3}%
\end{figure}%
\begin{par}
\bigpar%
The simulations in Figure \ref{NumFig:Ex3} correspond to
\begin{align}
\label{data.Ex3}%
\ga\in\{0.1,\,3,\,12,\,30,\,60,\,100\},\qquad%
k=0.1,\qquad\Psi^{\prime\prime}\at{x}=\exp\bat{-\max{\{x,\,0\}}^2},
\end{align}
and illustrate how the wave trains depend on $\ga$. The right
picture shows the \emph{traces} of the wave trains, that means the
closed curves
\begin{align*}
\phase\mapsto\bpair{X\at\phase}{V\at\phase},
\end{align*}
whose diameters increase with $\ga$. Surprisingly, we find a
\emph{nested family} of curves: The traces for different values of
$\ga$ do not intersect, but all traces for $\tilde{\ga}\leq\ga$ fill
out the interior of the trace for $\ga$. This observation indicates
the existence of a `hidden structure' for the nonlinear
advance-delay differential equation \eqref{Intro:TW.Eqn}. In
particular, there must be an equivalent planar Hamiltonian system
such that the traces coincide with the level sets of the
Hamiltonian. We cannot prove that the wave trains for fixed $k$ and
increasing $\ga$ give rise to a nested family of closed curves but
mention that a similar phenomenon can be observed for wave trains in
FPU chains, see \cite{Her09,HR08b}.
\end{par}
%
%
%
%
%
%
%
%
%
%
\providecommand{\bysame}{\leavevmode\hbox to3em{\hrulefill}\thinspace}
\providecommand{\MR}{\relax\ifhmode\unskip\space\fi MR }
\providecommand{\MRhref}[2]{%
  \href{http://www.ams.org/mathscinet-getitem?mr=#1}{#2}
}
\providecommand{\href}[2]{#2}

%
%

\begin{thebibliography}{DEFW93}

\bibitem[BK04]{BK04}
O.~M. Braun and Y.~S. Kivshar, \emph{The {F}renkel-{K}ontorova model:
  {C}oncepts, {M}ethods, and {A}pplications}, Theoretical and Mathematical
  Physics, Springer, 2004.

\bibitem[BZ06]{BZh06}
P.~W. Bates and Ch. Zhang, \emph{Traveling pulses for the {K}lein-{G}ordon
  equation on a lattice or continuum with long-range interaction}, Discr. Cont.
  Dynam. Systems Ser. A \textbf{16} (2006), no.~1, 235--252.

\bibitem[DEFW93]{DEFW93}
D.~B. Duncan, J.~C. Eilbeck, H.~Feddersen, and J.~A.~D. Wattis, \emph{Solitons
  on lattices}, Physica D \textbf{68} (1993), 1--11.

\bibitem[DPB93]{DPB93}
Th. Dauxois, M.~Peyrard, and A.~R. Bishop, \emph{Dynamics and thermodynamics of
  a nonlinear model for {DNA} denaturation}, Phys. Rev. E \textbf{47} (1993),
  no.~1, 684--695.

\bibitem[Her08]{Her09}
M.~Herrmann, \emph{Unimodal wave trains and solitons in convex {FPU} chains},
  arXiv:0901.3736, 2008.

\bibitem[HR08]{HR08b}
M.~Herrmann and J.~Rademacher, \emph{Heteroclinic travelling waves in convex
  {FPU}-type chains}, arXiv:0812.1712, 2008.

\bibitem[IJ05]{IJ05}
G.~Iooss and G.~James, \emph{Localized waves in nonlinear oscillator chains},
  Chaos \textbf{15} (2005), no.~1, 015113:1--15.

\bibitem[IK00]{IK00}
G.~Iooss and K.~Kirchg\"assner, \emph{Travelling waves in a chain of coupled
  nonlinear oscillators}, Comm. Math. Phys. \textbf{211} (2000), no.~2,
  439--464.

\bibitem[IP06]{IP06}
G.~Iooss and D.~E. Pelinovsky, \emph{Normal form for travelling kinks in
  discrete {K}lein {G}ordon lattices}, Physica D \textbf{616} (2006), no.~2,
  327--345.

\bibitem[Kat05]{Kat05}
Guy Katriel, \emph{Existence of travelling waves in discrete {S}ine-{G}ordon
  rings}, SIAM J. Math. Anal. \textbf{36} (2005), no.~5, 1434--1443.

\bibitem[KZ08a]{KZ08b}
C.~F. Kreiner and J.~Zimmer, \emph{Heteroclinic travelling waves for the
  lattice {S}ine-{G}ordon equation with linear pair interaction}, BICS preprint
  1/08, 2008.

\bibitem[KZ08b]{KZ08a}
\bysame, \emph{Travelling wave solutions for the discrete {S}ine-{G}ordon
  equation with nonlinear pair interaction}, BICS preprint 3/08, 2008.

\bibitem[MJKA02]{MJKS02}
A.~M. Morgante, M.~Johansson, G.~Kopidakis, and S.~Aubry, \emph{Standing wave
  instabilities in a chain of nonlinear coupled oscillators}, Phys. D
  \textbf{162} (2002), no.~1/2, 53--94.

\bibitem[Pan05]{Pan05}
A.~Pankov, \emph{Traveling waves and periodic oscillations in
  {F}ermi-{P}asta-{U}lam lattices}, Imperial College Press, London, 2005.

\end{thebibliography}
\end{document}